\documentclass[12pt]{article}
\usepackage{graphicx,cite,epsfig}
\newcommand{\beq}{\begin{equation}}
\newcommand{\eeq}{\end{equation}}
\newcommand{\barr}{\begin{array}}
\newcommand{\earr}{\end{array}}
\newcommand{\etal}{{\em et al.}}
\def\gev{\: \rm GeV} 
 
\newcommand{\dis}{\displaystyle}

\def\hr{\hookrightarrow}
\def\hl{\stackrel{\lambda}{\hookrightarrow}}
\def\tl{\stackrel{\lambda}{\rightarrow}}
\def\l{\lambda}
\def\e{\ell}
\def\snu{\widetilde{\nu}}

\def\stau{\widetilde{\tau}}

\def\slep{\widetilde{\ell}}
\def\N0{\widetilde{\chi}^0}
\def\Cp{\widetilde{\chi}^+}

\def\Cpm{\widetilde{\chi}^\pm}

\def\rp{R\!\!\!/ _p}
\catcode`@=11 
\def \gsim{\mathrel{\mathpalette\@versim>}}
\def \lsim{\mathrel{\mathpalette\@versim<}}
\def \@versim#1#2{\lower0.4ex\vbox{\baselineskip\z@skip\lineskip\z@skip
     \lineskiplimit\z@\ialign{$\m@th#1\hfil##\hfil$%
     \crcr#2\crcr\sim\crcr}}}
\catcode`@=12 
\begin{document}
\setcounter{page}{0}
\thispagestyle{empty}
\begin{flushright}
hep-ph/0412411 \\
\today
\end{flushright}
\begin{center}
{\Large\sc 
Tagging Sneutrino Resonances at a Linear Collider with Associated Photons} \\
\vspace*{0.2in}
{\large\sl Debajyoti Choudhury$^a$, Santosh Kumar Rai$^b$ {\rm and} 
Sreerup Raychaudhuri$^b$} \\
\vspace*{0.3in}
$^a$Department of Physics \& Astrophysics, University of Delhi, Delhi 110 007, 
India, \\ Electronic address: {\sf debchou@physics.du.ac.in} \\
$^b$Department of Physics, Indian Institute of Technology, Kanpur 208 016, 
India. \\ Electronic address: {\sf skrai@iitk.ac.in, sreerup@iitk.ac.in} \\
\vspace*{1.3in}
{\Large\bf ABSTRACT}
\end{center}
\begin{quotation} \noindent \sl
Sneutrino resonances at a high energy linear $e^+e^-$ collider may be
one of the clearest signals of supersymmetry without $R$-parity,
especially when the $R$-parity-violating coupling is too small to
produce observable excesses in four-fermion processes. However, there
is no guarantee that the sneutrino pole will lie anywhere near the
machine energy. We show that associated photon production induces the
necessary energy spread, and that the resonance then leaves a clear
imprint in the photon spectrum. It follows that tagging of a hard
mono-energetic photon for a variety of possible final states provides
a realistic method of separating sneutrino resonance signals from the
Standard Model backgrounds.
\end{quotation} \rm\normalsize

\vfill
\newpage
\section{Introduction}

\noindent 
In many models of supersymmetry,
$R$-parity\cite{Fayet:1977yc,Farrar:1978xj}, defined as $R = (-)^{L +
3B + 2S}$, where $L, B$ and $S$ stand, respectively, for the lepton
number, baryon number and spin of a particle, is often assumed to be
conserved. A conservation law of this kind is clearly tantamount to
separate conservation\footnote{Strictly speaking, this is true only 
if we restrict ourselves to renormalizable superpotentials.}
of both the global $U(1)$ quantum numbers $L$
and $B$. This idea, originally
introduced to combat fast proton decay, has since been shown to be
necessary only in part.  It is enough to conserve either lepton number
$L$ or baryon number $B$ -- and not both -- to have the requisite
stability for protons. Furthermore, there does not exist any other  
overriding theoretical motivation for imposing this symmetry. 
In fact, it has been argued~\cite{Ibanez:1992pr} that 
stability of the proton is better ensured by imposing a generalized 
baryon parity (a $Z_3$ symmetry) instead. Unlike $R$-parity, the latter 
also serves to eliminate dimension-5 operators that 
could potentially have led to proton decay. This 
has the added advantage that non-zero $\rp$ 
couplings provide a means of generating the small neutrino 
masses and large mixings that the 
neutrino oscillation experiments seem to call for. 

\bigskip\noindent In this paper, we assume that baryon number is
conserved and concentrate on $R$-parity violation through lepton
number-violating operators of the so-called $LL\bar{E}$ form. The
relevant term in the superpotential can be written as
\begin{equation} 
{\cal W}_{LL\bar{E}} = \l_{ijk} \epsilon_{ab} \hat{L}_i^a
\hat{L}_j^b \hat{E}_k  \ , \qquad i, j = 1\dots 3
\end{equation} 
where $\hat{L}_i \equiv (\hat{\nu}_{Li}, \hat{\ell}_{Li})^T$ and $\hat E_i$ 
are the $SU(2)$-doublet and singlet superfields respectively 
whereas $\epsilon_{ab}$ is the unit antisymmetric tensor.
Clearly, the coupling constants  $\l_{ijk}$ are antisymmetric 
under the exchange of the first two indices; 
the 9 such independent 
couplings are usually labelled keeping
$i>j$. In the above it is assumed that the chiral structure of the
Standard Model (SM) holds and any interaction terms with right-handed
neutrinos (or corresponding superfields) would be strongly suppressed
by the tiny neutrino masses. Written in terms of the component fields, 
the above superpotential leads to the interaction Lagrangian
\beq 
\barr{rcl}
{\cal L}_\l &
= & \dis \l_{ijk} \bigg[
   \snu^i     \bar{\e}_R^k           \e_L^j
+  \slep_L^j     \bar{\e}_R^k           \nu_L^i
+ (\slep_R^k)^*  \overline{(\nu_L^i)^c} \e_L^j
+ (\snu^i)^*  \bar{\e}_L^j           \e_R^k
+ (\slep_L^j)^*  \bar{\nu}_L^i         {\e}_L^k
+  \slep_R^k     \bar{\e}_L^j          (\nu_L^i)^c 
\\
& & \dis \hspace*{0.3in}
-  \snu^j     \bar{\e}_R^k           \e_L^i
-  \slep_L^i     \bar{\e}_R^k           \nu_L^j
- (\slep_R^k)^*  \overline{(\nu_L^j)^c} \e_L^i
- (\snu^j)^*  \bar{\e}_L^i           \e_R^k
- (\slep_L^i)^*  \bar{\nu}_L^j         {\e}_L^k
-  \slep_R^k     \bar{\e}_L^i          (\nu_L^j)^c
\bigg] \ .
\earr
     \label{gen_Lag}
\eeq

\bigskip\noindent
Just like the usual Yukawa couplings, the magnitude of the couplings
$\l_{ijk}$ are entirely arbitrary, and are restricted only from
phenomenological considerations.  The preservation of a GUT-generated
$B-L$ asymmetry, for example, necessitates the preservation of at
least one of the individual lepton numbers over cosmological time
scales~\cite{Dreiner:vm}.  Similarly, the failure of various collider
experiments~\cite{nosusy,HERA} to find any evidence of supersymmetry has
implied constraints in the parameter space.  Even if superpartners
were too heavy to be produced directly, strong bounds on these
couplings may still be deduced from the remarkable agreement between
low energy observables and the SM predictions.  These include, for
example, meson decay widths~\cite{Barger:1989rk,Bhattacharyya:1995pq},
neutrino masses~\cite{neutrino_mass,Bhattacharyya:1995pq}, rates for
neutrinoless double beta decay~\cite{bb0nu}, etc.  The bounds
generally scale with the sfermion mass and, for $m_{\tilde f} = 100
\gev$, they range from $\sim 0.02$ to $0.8$~\cite{rplimits}. In view 
of such constraints, strategies for collider signals for
$R_p$-violating supersymmetry are often designed for 
scenarios wherein the production of the
superparticles is dominated by gauge couplings and the leading role of
$R_p$-violation is in the decay of the lightest supersymmetric
particle~\cite{lspdecay}. Clearly, such studies would be insensitive to the
exact size of the $R_p$-violating coupling as long as it is large
enough to make the decay length of the LSP 
undetectable\footnote{If any of the $R_p$-violating 
                couplings is $> 10^{-6}$ or so, then the
                LSP will decay within the 
                detector~\protect\cite{Dawson:1985vr}.}. 
In contrast to this,  processes directly sensitive to the size of such
couplings would include ($i$) production of sparticles 
through them~\cite{Moreau:2000bs,HERA_expl,DCSR}, ($ii$) the 
decays of sparticles through them~\cite{Ghosh:1997bm,hanmargo} and
($iii$) modification of SM amplitudes through exchanges 
of virtual sparticles~\cite{Bhattacharyya:1994yc,
Kal, Hewett:1997ce, Ghosh:1997bm,Hikasa:1999wy,CCQR,DCSR}. 
An accurate measurement of such cross-sections can, apart from leading 
to the discovery of supersymmetry, also serve as a means of measuring 
the size of such couplings. In this paper,
we concentrate on one such example, namely single sneutrino production 
in association with a hard photon.

\bigskip\noindent
As we shall see presently, the terms relevant for our discussion are
the first and fourth ones on both first and second lines of 
eqn.(\ref{gen_Lag}), with $j = k
=1$ on the first line and $i = k = 1$ on the second. Isolating these
leads to the specific interactions
\beq
\barr{rcl}
{\cal L}_\l 
& = & 
- 2\l_{1j1} \, \bigg[ \snu^j \, \bar{e}_R \, e_L + 
         (\snu_L^j)^* \, \bar{e}_L \, e_R \bigg] + \cdots
 \\[2ex] 
& = & 
- 2\l_{121} \, \bigg[ \snu_{\mu} \, \bar{e}_R \, e_L + 
          (\snu_{\mu})^* \, \bar{e}_L \, e_R \bigg] 
- 2\l_{131} \, \bigg[ \snu_{\tau} \, \bar{e}_R \, e_L + 
          (\snu_{\tau})^* \, \bar{e}_L \, e_R \bigg] + \cdots
\earr
\eeq
where the dots stand for the terms in eq.(\ref{gen_Lag}) that 
are irrelevant to the present discussion. 
It is then a simple matter to read off the Feynman rules for the vertices 
$$
e^+e^-\snu_\mu \ , \qquad 
e^+e^-\snu_\mu^* \ , \qquad 
e^+e^-\snu_\tau \ , \qquad
e^+e^-\snu_\tau^* \ .
$$
The presence of these vertices clearly leads to resonances in the processes
\cite{Feng:1996}
\begin{eqnarray}
e^+ ~+~ e^- & \longrightarrow & \snu_{\mu/\tau} (\snu_{\mu/\tau}^*)
\longrightarrow e^+ ~+~ e^-
\nonumber \\
& \hookrightarrow & \snu_{\mu/\tau} (\snu_{\mu/\tau}^*)
  \longrightarrow \nu_{\mu/\tau} (\overline{\nu}_{\mu/\tau}) ~+~ \N0_{1/2/3/4}
\nonumber \\
& \hookrightarrow & \snu_{\mu/\tau} (\snu_{\mu/\tau}^*)
\longrightarrow \mu^\mp/\tau^\mp ~+~ \Cpm_{1/2}
\nonumber \ ,
\end{eqnarray}
the first of which resembles Bhabha scattering in QED or in the SM. As
in Bhabha scattering, the sneutrino exchange can occur in both $s$ and
$t$ channels.  For the other two it is simply an $s$-channel sneutrino
exchange. It is also implicit that only those processes among the
above will occur which are kinematically allowed, i.e. if the higher
neutralino and chargino states are heavier than the sneutrino, the
corresponding process will occur off-shell, with strong propagator
suppression of the corresponding cross-sections.

\bigskip\noindent 
The sneutrino decay width, which is a simple matter
to compute, never rises above 3--4 GeV, which means that at a collider
with several hundred GeVs of energy, we can apply the {\it narrow-width
approximation} with impunity. In this work, therefore, we solely
consider on-shell production of sneutrinos (of muonic or tauonic
flavour). It is also worth mentioning that, in line with most of the
literature on $R$-parity violation, we consider only one non-vanishing
(or dominant) $\l$-coupling, 
for the simultaneous presence of more than 
        one $\rp$ coupling could potentially lead to flavour-changing 
        neutral currents and hence is subject to rather 
stringent constraints~\cite{fcnc}. Though apparently unnatural, this is not
more so than the pattern of Yukawa couplings in the SM.

\bigskip\noindent The principal issue on which this work hinges is the
fact that a high-energy $e^+e^-$ collider is likely to run at just a
single (or a few fixed) centre-of-mass energies $\sqrt{s}$. Given our
present lack of knowledge of sneutrino masses (or even of the
existence of sneutrinos) it is highly unlikely that for these
pre-determined machine energies we can have $\sqrt{s} \approx
m_{\snu}$. If, indeed, $m_{\snu} < \sqrt{s}$, then the cross-section
for $e^+ ~+~ e^- \longrightarrow \snu_{\mu/\tau} (\snu_{\mu/\tau}^*)$
will be strongly propagator-suppressed. However, if we consider the
process $e^+ ~+~ e^- \longrightarrow \gamma + \snu_{\mu/\tau}
(\snu_{\mu/\tau}^*)$, then, for some of the events, the photon may
carry-off just enough energy for the remaining $e^+e^-$ system to
excite the sneutrino resonance.  To borrow from a much used 
terminology, we are essentially considering a ``radiative return
to the sneutrino''. A similar method of detecting massive graviton 
resonances has been discussed in Ref.~\cite{Rai:2003}.
With processes of interest being 
of the form 
\beq
\barr{rcl} e^+ ~+~ e^- &
\longrightarrow & \gamma ~+~ \snu_{\mu/\tau} (\snu_{\mu/\tau}^*) 
\longrightarrow
\gamma ~+~ e^+ ~+~ e^- 
 \\ 
& \hookrightarrow & \gamma ~+~ \snu_{\mu/\tau}
(\snu_{\mu/\tau}^*) \longrightarrow \gamma ~+~ \nu_{\mu/\tau}
(\overline{\nu}_{\mu/\tau}) ~+~ \N0_{1/2/3/4}
  \\
 & \hookrightarrow & \gamma ~+~ \snu_{\mu/\tau} (\snu_{\mu/\tau}^*)
 \longrightarrow \gamma ~+~ \mu^\mp/\tau^\mp ~+~ \Cpm_{1/2} \ ,
\earr
\eeq
the application of the narrow-width approximation
ensures an almost monochromatic photon of energy
\begin{equation}
E_\gamma = \frac{s - m_{\snu}^2}{2\sqrt{s}} \ .
\end{equation}
This, potentially, would stand out
against the continuum spectrum arising from the Standard Model
background. Since the sneutrino $\snu_{\mu/\tau}$ can have a variety
of decay channels, we can simply tag on a hard isolated photon
associated with any of these decay channels and look for a line
spectrum superposed on the continuum background. This will lead, as
our discussion will show, to clear signals of sneutrino
production. Moreover, the $R$-parity-violating decays of the sneutrino
will set up multi-lepton final states (with associated photons) which
will have little or no Standard Model backgrounds worth
considering. For such states a mono-energetic photon will clinch the
issue of sneutrino production. Our work establishes, therefore, that
at a linear collider, $R$-parity-violating supersymmetry may be
detected early through an associated photon, perhaps even before the
conventional supersymmetry searches have collected enough statistics.

\bigskip\noindent It is worth noting that the usual signature for
$R$-parity-violating supersymmetry at an $e^+e^-$ collider is through
four-fermion processes of the form (for $LL\bar E$ operators) $e^+e^-
\to e^+e^-$, $e^+e^- \to \mu^+\mu^-$ and $e^+e^- \to \tau^+\tau^-$,
where the principal contribution is through $t$-channel sneutrino
exchange. A simple consideration of the excess (over the SM)
cross-section\cite{DCSR} leads to a discovery limit of about $\l_{1j1}
= 0.04$ for the lepton number-violating coupling responsible for the
signal, for $m_{\snu} \lsim 200 \gev$. Our work, is,
therefore, principally concerned with signal for $R$-parity violation
when the coupling is $\l_{1j1} \lsim 0.04$, but the sneutrino is
light enough to be produced as a resonance in $e^+e^-$ collisions at
the machine energies of 500~GeV and 1~TeV. In fact, one major
advantage of studying resonant sneutrinos is that the parameter space
of the model can be explored almost upto $m_{\snu} = \sqrt{s}$, except
for a small reduction due to kinematic cuts on the final states
observed. By contrast, slepton or sneutrino pair production has the
potential to explore only $m_{\snu} \leq
\sqrt{s}/2$.

\bigskip\noindent This paper is organised as follows. In the next two
sections we discuss, successively, the production cross-section and
different decay channels of the two sneutrinos which are under
investigation. Section 4 is devoted to a discussion of backgrounds and
possible strategies to isolate the signal. In section 4, we discuss how a
sneutrino resonsnce could be distinguished from other possible 
new physics effects. And finally, section 6 contains our
conclusions and some general comments.

\section{Sneutrino production with associated photons}

\begin{figure}[htb] 
\begin{center} 
\vspace*{2.6in} 
\relax\noindent\hskip -4.0in\relax{\includegraphics{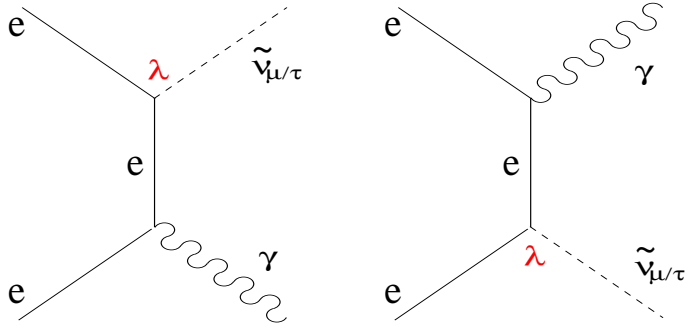}} 
\end{center} 
\vspace*{-1.4in}
\caption{\footnotesize\it Feynman diagrams for sneutrino
production with associated photons at a linear collider.}
   \label{fig:feynman}
\end{figure}

\bigskip\noindent The specific reaction on which we focus in this paper
is the associated photon process
$$
e^+ ~+~ e^- \longrightarrow \gamma ~+~ \snu_{\mu/\tau} (\snu_{\mu/\tau}^*)
$$ 
illustrated in Figure~\ref{fig:feynman}. The squared and spin-averaged matrix
element for this is, then
\begin{equation}
\overline{|{\cal M}|^2} = 8\pi\alpha~\l_{1j1}^2 \frac{s^2 + \widetilde{m}_j^4}{tu}
~\theta(s - \widetilde{m}_j^2)
    \label{mesq}
\end{equation}
where $\widetilde{m}_j$ is the mass of the muonic ($j=2$) or tauonic
($j=3$) sneutrino. 
\bigskip 
The cross-sections for production of the sneutrino
and its anti-particle are identical; if we do not distinguish between
the signals for these, the effective cross-section must be multiplied
by a factor of 2.  The collinear singularity in eqn.(\ref{mesq}), so 
characteristic of  massless electrons and photons, is automatically 
taken care of once one imposes restrictions on the phase space
commensurate with the detector acceptances. In the rest of the 
analysis, we shall require the photon to be sufficiently hard 
and transverse, namely
\beq
\barr{lcl}
{\rm pseudorapidity}: & \qquad& 
    \left| \eta_\gamma \right| < \eta_\gamma^{(max)} = 2.0 \ , \\
{\rm transverse \ momentum}: & & p_{T\gamma} > p_{T\gamma}^{(min)} = 20 \gev \ .\earr
    \label{photon_cuts}
\eeq
Integrating eqn.(\ref{mesq}) leads to a production cross-section of the form
\begin{equation}
 \sigma(x_j) = \frac{2 \alpha \lambda_{1j1}^2}{s} \;
           \frac{1 + x_j^2}{1 - x_j} \; \theta(1 - x_j)
\; \times ~{\rm min}\left[ \,\eta_\gamma^{(max)},
\; \log \frac{1 - x_j - \sqrt{(1 - x_j)^2 - 4 x_T^2}}{2 x_T} \, \right]
\label{csecn}
\end{equation}
where $x_j = {\widetilde{m}_j^2}/{s}$ and $x_T = p_{T\gamma}^{(min)}/\sqrt{s}$.

\begin{figure}[htb]
\centerline{\includegraphics[height=3in]{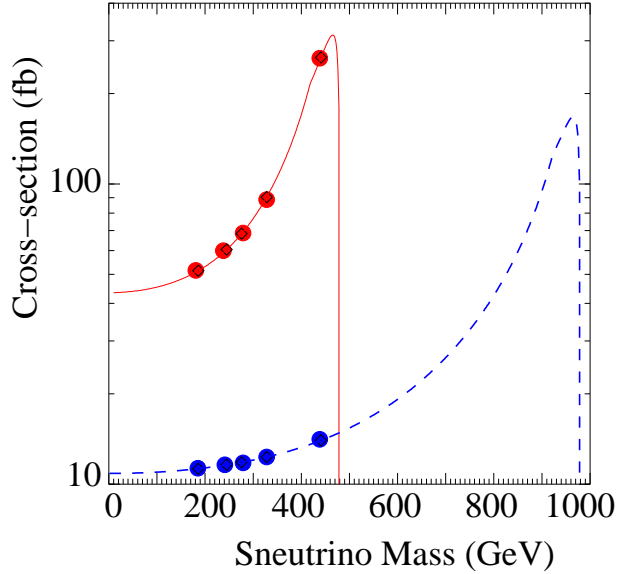}}
\caption{\footnotesize\it Cross sections for
sneutrino production with associated photons at a linear collider for
$\l_{1j1} = 0.03$. Solid red (dashed blue) lines correspond to a 
500~GeV (1~TeV)
centre-of-mass energy. The cuts of eqns.(\protect\ref{photon_cuts}) 
have been imposed. The points marked with bullets are for a
$\snu_\mu$ resonance at the Snowmass MSugra points 1$a$, 1$b$, 3, 4,
and 5. At the point 2, the sneutrino is beyond the kinematic reach of
the linear collider. If sneutrinos are not distinguished from
anti-sneutrinos, the cross-section(s) would be doubled.}
\label{fig:prodn}
\end{figure}

\bigskip\noindent
The cross-section in Eqn. (\ref{csecn}) is plotted in
Figure~\ref{fig:prodn} as a function of $\widetilde{m}_j$ and with
$\l_{1j1} = 0.03$ for a linear collider running at ($a$)~500~GeV and
($b$)~1~TeV. Several comments are in order:
\begin{itemize}
\item Contrary to naive expectations, the cross section rises with 
the sneutrino mass (until nearly the kinematic limit). This is occasioned 
by the fact that a sneutrino mass closer to the centre of mass energy 
implies that the the photon needs to carry off less energy, thereby 
facilitating the radiative return to the sneutrino. 
\item A consequence of the above is the fact that, below the kinematic limit,
the cross sections at a 500 GeV collider are larger than those at a 
1 TeV machine.
\item The steep fall in the cross-section when the kinematic limit is 
approached occurs just a little before the actual threshold $m_{\snu} 
= \sqrt{s}/2$. This is simply a consequence of our having
demanded a nonzero $(p_{T\gamma})^{(min)}$. 
\item It is also interesting to note the very slight kink in the graph(s)
a little before the fall-off. This is another artefact of our cuts. Below this
mass it is the restriction on $\eta_\gamma$ that is mainly operative, while
above this mass, it is the cut on $p_{T\gamma}$ that takes over. This
is a discrete transition, causing the slight kink as mentioned.
\item As the graph shows, we obtain cross-sections typically in the 
range 50 fb--250 fb. 
At a linear collider with around 500 fb$^{-1}$ of integrated
luminosity, this amounts to the production of a very large number of
sneutrinos along with an associated monochromatic photon. 
Thus, even if $\lambda_{1j1}$ were to be 
smaller by an order of magnitude, we would still have a fairly 
large number of such distinctive events.
 It is clear, therefore, that if the sneutrino is kinematically
accessible to a linear collider, low statistics will not be the major
hurdle in their detection.
\end{itemize}

\bigskip\noindent  
Although there is no strict 
restriction on the mass spectrum of $R$-parity-violating models
(except for weak experimental bounds from LEP-2 and the Tevatron), it is 
useful, for the purpose of easy comparison, 
to focus on the MSugra spectrum and, specifically, on
the six representative points chosen at the 2001 Snowmass
conference. The latter are described by

\bigskip
\begin{tabular}{clllll}
\bf 1a : & $M_0 = ~100$~GeV, &  $m_{1/2} = 250$~GeV,&  $A_0 = -100$~GeV,
& $\tan\beta = 10$,&  $\mu >0$   \\
\bf 1b : & $M_0 = ~200$~GeV, &  $m_{1/2} = 400$~GeV,&  $A_0 = 0$,
&  $\tan\beta = 30$,&  $\mu >0$   \\
\bf ~2 : & $M_0 = 1450$~GeV,&  $m_{1/2} = 300$~GeV,&  $A_0 = 0$, 
& $\tan\beta = 10$, & $\mu >0$   \\
\bf ~3 : & $M_0 =  90$~GeV, & $m_{1/2} = 400$~GeV,&  $A_0 = 0$, 
& $\tan\beta = 10$, & $\mu >0$   \\
\bf ~4 : & $M_0 = ~400$~GeV, &  $m_{1/2} = 300$~GeV,&  $A_0 = 0$, 
& $\tan\beta = 50$, & $\mu >0$   \\
\bf ~5 : & $M_0 = ~150$~GeV, &  $m_{1/2} = 300$~GeV,&  $A_0 = -1$~TeV, 
& $\tan\beta = ~~5$, & $\mu >0$ \\
\end{tabular}

\bigskip\noindent  
At most of these Snowmass points, the muonic and tauonic sneutrinos 
are almost degenerate (see Table.~\ref{table:spectrum}).
In Figure~\ref{fig:prodn}, five of the six mSugra points (1a, 1b, 3--5)
have been marked with bullets ($\bullet$). The Snowmass point numbered 2
leads to a sneutrino mass of 1.45 TeV which is clearly out of the
kinematic range of a 500~GeV or even a 1~TeV linear collider.
Note also that in each of these cases, the lightest neutralino is the lightest
supersymmetric particle 
(LSP)\footnote{While this is a requirement for a $R$-parity conserving 
    model to be phenomenologically viable, it clearly is not so 
    in the event of a broken  $R$-parity.}
This is quite clear from Table~1, where we have
listed the sneutrino masses and the masses of their daughters in possible decay
channels.
\begin{table}[htb]
\small
\begin{center} $$
\begin{array}{|c||c|c|c|c|c|c|c|} \hline
{\rm Point} & \snu_\mu & \snu_\tau & \stau_1 & 
\stau_2 & \N0_1 & \N0_2 & \Cpm_1 \\ \hline\hline
1a & 186 & 185 & 133 & 206 & ~96 & 177 & 176 \\
1b & 328 & 317 & 196 & 344 & 160 & 299 & 299 \\
~2 & 1454& 1448& 1439& 1450& ~80 & 135 & 104 \\
~3 & 276 & 275 & 171 & 289 & 161 & 297 & 297 \\
~4 & 441 & 389 & 268 & 415 & 119 & 218 & 218 \\
~5 & 245 & 242 & 181 & 258 & 120 & 226 & 226 \\ 
\hline
\end{array} $$
\end{center} \normalsize
\caption{\footnotesize\sl Relevant parts of the MSugra spectrum
for the six Snowmass points. All masses are in GeV, rounded off to the nearest
whole number.}
    \label{table:spectrum}
\end{table}

\bigskip\noindent
The major hurdle in detection of sneutrinos will, of
course, be isolation of the signal from the substantial Standard Model
backgrounds, since massive sneutrinos can decay in a variety of
channels, each with characteristic signatures. As there is a
$R$-parity-violating $\l$ coupling, we should expect several types of
hadronically-quiet multi-lepton signals. These are discussed in the
section which follows.
\section{Decays of the Sneutrino}

\bigskip\noindent Some of the sneutrino decays have already figured in
the discussion of resonant processes. However, in $R$-parity-violating
models, decays of the sneutrino are rather complex.
Two distinct scenarios are identifiable though:

\begin{itemize}
\item {\it Small-$\l$ limit}: When the $R$-parity-violating coupling
$\l_{1j1}$ is much smaller than the gauge couplings, the sneutrino
decays principally through normal, $R$-parity-conserving, channels. In
this case, $R$-parity violation manifests itself only in the decays of
the neutralino LSP. Note, however, that unless $\lambda_{1j1} \lsim 10^{-5}$ 
the LSP also decays almost at the interaction
point into an invisible neutrino and a lepton pair (not necessarily of
the same flavour). The final signal will, therefore, include the
daughters of sneutrino decays as well as those arising from LSP
decays. There are several such possibilities, depending on the mass
spectrum, and hence each point in the parameter space has to be
considered separately.
\item {\it Large-$\l$ limit}: When the $R$-parity-violating coupling
$\l_{1j1}$ is comparable to the gauge
couplings,  the sneutrino will have a substantial decay width into an
$e^+e^-$ pair. In fact, this may even become the dominant decay mode.
\end{itemize}

\begin{table}[htb]
\scriptsize
\begin{minipage}{3.2in}
$$
\begin{array}{|c|ccl|ccl|}\hline
& \multicolumn{3}{c|}{\bf\l_{121}}  & \multicolumn{3}{c|}{\bf\l_{131}}  \\ 
\hline\hline
       & \snu_\mu &\tl& e^-   \,+\, e^+                    & \snu_\tau &\tl & e^-    \,+\, e^+                      \\
       &          &\hr& \nu_\mu \,+\, \N0_{1,2}               &           &\hr &\nu_\tau  \,+\, \N0_{1,2}                \\
       &          &\hr& \mu^- \,+\, \Cp_1                   &           &\hr & \tau^- \,+\, \Cp_1                    \\
       & \N0_{1,2}&\tl& e^-   \,+\, e^+     \,+\,\bar{\nu}_\mu & \N0_{1,2} &\tl & e^-    \,+\, e^+     \,+\,\bar{\nu}_\tau \\
{\bf 1a} &        &\hl& e^-   \,+\, \mu^+   \,+\,\bar{\nu}_e   &           &\hl & e^-    \,+\, \tau^+  \,+\,\bar{\nu}_e    \\
       & \N0_2    &\to& \tau^-  \,+\, \stau^+                 & \N0_2     &\to & \tau^- \,+\, \stau^+                  \\
       & \Cp_1    &\to& \stau^+ \,+\, \nu_\tau                & \Cp_1     &\to & \stau^+  \,+\, \nu_\tau                 \\
       & \stau^+  &\to& \N0_1 \,+\, \tau^+                  & \stau^+   &\to & \N0_1  \,+\, \tau^+                   \\ \hline\hline
       & \snu_\mu &\tl& e^-   \,+\, e^+                     & \snu_\tau &\tl & e^-    \,+\, e^+                      \\
       &          &\hr& \nu_\mu \,+\, \N0_{1,2}               &           &\hr &\nu_\tau  \,+\, \N0_{1,2}                \\
       &          &\hr& \mu^- \,+\, \Cp_1                   &           &\hr & \tau^- \,+\, \Cp_1                    \\
       &          &   &                                  &           &\hr & \stau^-  \,+\, W^+                      \\
       & \N0_{1,2}&\tl& e^-   \,+\, e^+     \,+\,\bar{\nu}_\mu & \N0_{1,2} &\tl & e^-    \,+\, e^+     \,+\,\bar{\nu}_\tau \\
{\bf 1b} &        &\hl& e^-   \,+\, \mu^+   \,+\,\bar{\nu}_e   &           &\hl & e^-    \,+\, \tau^+  \,+\,\bar{\nu}_e    \\
       & \N0_2    &\to& \tau^-  \,+\, \stau^+                 & \N0_2     &\to & \tau^- \,+\, \stau^+                  \\
       & \Cp_1    &\to& \stau^+ \,+\, \nu_\tau                & \Cp_1     &\to & \stau^+  \,+\, \nu_\tau                 \\
       &          &\hr& \N0_1 \,+\, W^+                     &           &\hr & \N0_1  \,+\, W^+                      \\
       & \stau^+  &\to& \N0_1 \,+\, \tau^+                  & \stau^+   &\to & \N0_1  \,+\, \tau^+                \\ \hline\hline
\end{array} $$ 
\vspace*{80pt}
\end{minipage}
\hfill
\begin{minipage}{3.2in}
\scriptsize
$$
\begin{array}{|c|ccc|ccc|}\hline
& \multicolumn{3}{c|}{\bf\l_{121}}           & \multicolumn{3}{c|}{\bf\l_{131}}  \\ 
\hline\hline
       & \snu_\mu &\tl& e^-   \,+\, e^+                     & \snu_\tau &\tl & e^-    \,+\, e^+                      \\
       &          &\hr& \nu_\mu \,+\, \N0_1                   &           &\hr &\nu_\tau  \,+\, \N0_1                    \\
{\bf 3} &         &   &                                  &           &\hr &\stau^- \,+\, W^+                      \\
       & \N0_1    &\tl& e^-   \,+\, e^+     \,+\,\bar{\nu}_\mu & \N0_1     &\tl & e^-    \,+\, e^+     \,+\,\bar{\nu}_\tau \\
       &          &\hl& e^-   \,+\, \mu^+   \,+\,\bar{\nu}_e   &           &\hl & e^-    \,+\, \tau^+  \,+\,\bar{\nu}_e    \\          
       &          &   &                                  & \stau^+   &\to & \N0_1  \,+\, \tau^-                   \\ \hline\hline
       & \snu_\mu &\tl& e^-   \,+\, e^+                     & \snu_\tau &\tl & e^-    \,+\, e^+                      \\
       &          &\hr& \nu_\mu \,+\, \N0_{1,2}               &           &\hr &\nu_\tau  \,+\, \N0_{1,2}                \\
       &          &\hr& \mu^- \,+\, \Cp_1                   &           &\hr & \tau^- \,+\, \Cp_1                    \\
       &          &   &                                  &           &\hr & \stau^-  \,+\, W^+                      \\
       & \N0_{1,2}&\tl& e^-   \,+\, e^+     \,+\,\bar{\nu}_\mu & \N0_{1,2} &\tl & e^-    \,+\, e^+     \,+\,\bar{\nu}_\tau \\
{\bf  4} &        &\hl& e^-   \,+\, \mu^+   \,+\,\bar{\nu}_e   &           &\hl & e^-    \,+\, \tau^+  \,+\,\bar{\nu}_e    \\
       & \N0_2    &\to& \N0_1 \,+\, Z                       & \N0_2     &\to & \N0_1  \,+\, Z                        \\
       & \Cp_1    &\to& \N0_1 \,+\, W^+                     & \Cp_1     &\to & \N0_1  \,+\, W^+                      \\
       &          &   &                                  & \stau^+   &\to & \N0_{1,2}\,+\, \tau^+                   \\       
       &          &   &                                  & \stau^+   &\to & \Cp_1  \,+\, \bar{\nu}_\tau           \\ \hline\hline
       & \snu_\mu &\tl& e^-   \,+\, e^+                     & \snu_\tau &\tl & e^-    \,+\, e^+                      \\
       &          &\hr& \nu_\mu \,+\, \N0_{1,2}               &           &\hr &\nu_\tau  \,+\, \N0_{1,2}                \\
       &          &\hr& \mu^- \,+\, \Cp_1                   &           &\hr & \tau^- \,+\, \Cp_1                    \\
       & \N0_{1,2}&\tl& e^-   \,+\, e^+     \,+\,\bar{\nu}_\mu & \N0_{1,2} &\tl & e^-    \,+\, e^+     \,+\,\bar{\nu}_\tau \\
{\bf 5} &         &\hl& e^-   \,+\, \mu^+   \,+\,\bar{\nu}_e   &           &\hl & e^-    \,+\, \tau^+  \,+\,\bar{\nu}_e    \\
       & \N0_2    &\to& \tau^-  \,+\, \stau^+                 & \N0_2     &\to & \tau^- \,+\, \stau^+                  \\
       & \Cp_1    &\to& \stau^+ \,+\, \nu_\tau                & \Cp_1     &\to & \stau^+  \,+\, \nu_\tau                 \\
       &          &\hr& \N0_1 \,+\, W^+                     &           &\hr & \N0_1  \,+\, W^+                      \\
       & \stau^+  &\to& \N0_1 \,+\, \tau^+                  & \stau^+   &\to & \N0_1  \,+\, \tau^+               \\ \hline\hline
\end{array} $$ 
\end{minipage}
\normalsize

\caption {\footnotesize\sl Principal decay modes of
the sneutrinos $\snu_\mu$ and $\snu_\tau$, including
$R$-parity-violating decays, at five of the six Snowmass 2001
points. We exclude point {\bf 2} because it predicts that
$\snu$-production would be kinematically disallowed at a 500 GeV or
even a 1~TeV collider.}
   \label{table:decays}
\end{table}

\bigskip\noindent 
While the small-$\l$ limit 
simplifies the decay analysis, it also leads to a suppression of the 
production cross-section, which is proportional to $\l_{1j1}^2$. We,
therefore,  focus on the intermediate case, 
namely\footnote{ This value is also consistent with the
bounds expected from fermion pair production at a linear
collider\cite{DCSR}.} 
$\l_{1j1} \lsim 0.03$, which while somewhat smaller than the gauge
couplings, still allows the sneutrino and LSP to decay almost at
the interaction point. With this assumption, the principal decay modes
of the sneutrino(s) are those listed in Table~2.

\bigskip\noindent The decay modes marked with a $\l$ in Table~2 occur
when the sneutrino decays purely through the $R$-parity-violating
coupling $\lambda_{1j1}$.  In the remaining modes, the sneutrino
decays through $R$-parity-conserving (gauge) couplings, with an LSP at
the final stage of the cascades. The LSP then undergoes a three-body
decay through the same $\lambda_{1j1}$ coupling, with exchange of
virtual sleptons. The final states are described without specific
leptonic charges, partly because it may be operationally difficult to
tag lepton charges, and more importantly, because the Majorana nature
of the neutralino enables it to decay to either charge of each
leptonic flavour. (This also means that the probability gets
multiplied by a factor of two for each neutralino when the
$CP$-conjugated process is also considered.) Cascade decays of higher
gaugino states to the LSP through a third (intermediate) gaugino state
are discounted as the branching ratios are relatively small. Final
state $W$ and $Z$ bosons, will, of course, decay into all possible
fermion pairs, according to the branching ratios, increasing the
number of possible combinations. Since we are interested only in {\it
hadronically-quiet} signals, we do not consider their (dominant)
decays to quarks, but focus on the leptonic decays 
only\footnote{In this work, we have not considered final states with 
    jets, not because they are not important for the detection of 
    sneutrinos, but simply because the leptonic final states are cleaner 
    and easier to analyse. It also means that we can use a parton-level 
    Monte Carlo event generator without much error.}. 
Even with this simplification,
we still have a large number of possible final states which can appear
together with an associated (hard) photon. These are listed in
Table~3, with the cross-sections for the five Snowmass points which
are kinematically accessible. The table has been constructed assuming
$\lambda_{1j1} = 0.03$ as explained above.  We have convoluted the
cross-sections in Table~2 with detection efficiencies $\eta_e \simeq
\eta_\mu \simeq 90\%$ and $\eta_\tau \simeq 80\%$, which are
consistent with the known LEP-2 efficiency factors and
likely to be bettered at the NLC.

\bigskip\noindent Table~3 shows that the 18 types of $R$-parity-violating 
signals resolve themselves into four classes. These are
\begin{enumerate}
\item photon plus dielectron; 
\item photon plus dielectron plus missing energy; 
\item photon plus dileptons of dissimilar flavour plus missing energy; 
\item photon plus four leptons plus missing energy. 
\end{enumerate}
\begin{table}[htb]
\footnotesize
\begin{center} $$
\begin{array}{|c|l|cc|cc|cc|cc|cc|} \hline
\bf{(A)} & {\rm Final~state}
& \multicolumn{2}{c|}{\bf 1a}  & \multicolumn{2}{c|}{\bf 1b} & \multicolumn{2}{c|}{\bf 3} &
 \multicolumn{2}{c|}{\bf 4} & \multicolumn{2}{c|}{\bf 5}    \\ \hline\hline
\lambda_{121} & \sqrt{s}~({\rm TeV})  & 0.5 & (1.0) & 0.5 & (1.0) & 0.5 & (1.0) & 0.5 & (1.0) & 0.5 & (1.0) \\ \hline\hline
 1 & \gamma+ee
&  559 &(123)   &  823 & (113)  & 1058 & (183)   &  469 & ( 21)  &  580 & (111)  \\ \hline
 2 & \gamma+ee\not{\!\!E}
& 1592 & (349)    & 2256 & (308)  &2236 & (387)    & 2601 & (115)  & 1667 & (320)   \\ \hline
 3 & \gamma+e\mu\not{\!\!E}
& 1592 & (349)    & 2256 & (308)  &2236 & (387)    & 2601 & (115)  & 1667 & (320)  \\ \hline
 4 & \gamma+ee\tau\tau\not{\!\!E}
& 33& (-)      & 88 & (12) & -  & (-) & - & (-)  & 42 & (-) \\ \hline
 5 & \gamma+e\mu\tau\tau\not{\!\!E}
& 33& (-)      & 88 & (12) & -  & (-) & - & (-)  & 42 & (-) \\ \hline
 6 & \gamma+eee\mu\not{\!\!E}
&  - & (-)   &       - & (-)  &  - & (-)    &   658 & (2.9)  & - & (-) \\ \hline
 7 & \gamma+ee\mu\mu\not{\!\!E}
&  - & (-)   &   11   & (-)  &  - & (-)    &  1316 & (58)  & 15 & (-) \\ \hline
 8 & \gamma+ee\mu\tau\not{\!\!E}
&112 & (24)     &  446 & (61)   & - & (-)    &   585 & (26)  &  194 & (37)  \\ \hline
 9 & \gamma+e\mu\mu\tau\not{\!\!E}
&112 & (24)     &  446 & (61)   &      - & (-)    &   585 & (26)  &  194 & (37)  \\ \hline
10 & \gamma+e\mu\mu\mu\not{\!\!E}
&       - & (-)   &      - & (-)   &      - & (-)    &   658 & (29)  &       - & (-)  \\ \hline
\end{array} $$
$$
\begin{array}{|c|l|cc|cc|cc|cc|cc|} \hline
\bf{(B)} & {\rm Final~state}
& \multicolumn{2}{c|}{\bf 1a}  & \multicolumn{2}{c|}{\bf 1b} & \multicolumn{2}{c|}{\bf 3} &
 \multicolumn{2}{c|}{\bf 4} & \multicolumn{2}{c|}{\bf 5}    \\ \hline\hline
\lambda_{131} & \sqrt{s}~({\rm TeV})  & 0.5 & (1.0) & 0.5 & (1.0) & 0.5 & (1.0) & 0.5 & (1.0) & 0.5 & (1.0) \\ \hline\hline
 1 & \gamma+ee
&  576 &(126)  &  547 &( 79)  &  949 &(165)  &  256 &( 22) &  521 &(101) \\ \hline
 2 & \gamma+ee\not{\!\!E}
& 1618 &(355)  & 1387 &(200)  &1987 &(345)   & 1157 &(100) &1464 &(283) \\ \hline
 3 & \gamma+e\tau\not{\!\!E}
& 1438 &(316)   & 1233 &(178)  &1766 &(307)   & 1029 &( 89) &1301 &(251) \\ \hline
 4 & \gamma+ee\tau\tau\not{\!\!E}
& 108 &(24) & 350 &( 50)   &  41 &(-)    &  527 &(46)  & 155 &( 30) \\ \hline
 5 & \gamma+e\tau\tau\tau\not{\!\!E}
& 96  &(21)  & 217 &( 31)   &   18 &(-)   &  234 &(20)  & 134 & (26)    \\ \hline
 6 & \gamma+eee\tau\not{\!\!E}
&   - & (-)     & 119 &( 17)   &  23  &(-)   &  297 &(26) &    - & (-)    \\ \hline
 7 & \gamma+ee\mu\tau\not{\!\!E}
&   - & (-)     & 119 &( 17)   &  23  &(-)   &  297 &(26) &    - & (-)    \\ \hline
 8 & \gamma+e\mu\tau\tau\not{\!\!E}
&   - & (-)     &  106 &(15) &    21&(-)   &  264 &(23) &    -  &(-) \\ \hline
\end{array} $$
\end{center}

\caption{{\footnotesize\sl Number of events for luminosity
${\mathcal L}=100~fb^{-1}$ for different final states arising from
sneutrino decay cascades in {\rm (A)} $e^+ ~+~ e^- \longrightarrow
\gamma ~+~ \snu_{\mu} (\snu_{\mu}^*)$ and {\rm (B)} $e^+ ~+~ e^-
\longrightarrow \gamma ~+~ \snu_{\tau} (\snu_{\tau}^*)$ at a
~$500~~(1000)$~GeV $e^+~e^-$ linear collider with unpolarized
beams. Columns correspond to the Snowmass points (except the
kinematically disallowed {\bf 2} point). Entries marked with a dash
(--) indicate less than 10 events.  Detection efficiencies are
(crudely) included in the cross-section figures.} } 
    \label{table:finalstate}
\end{table}
\normalsize

\bigskip\noindent The first kind arises from the direct
$R$-parity-violating decay of the sneutrino and would have a large SM
background from radiative Bhabha scattering. The second and third ones
are obviously reproduced by $WW$-production. The last type arises from
higher-order effects in the SM and has very little background. Thus
each signal requires to be discussed separately and specific cuts and
isolation techniques need to be applied in each case. Of course, the
trigger will still be a (approximately) monochromatic photon, which
results from its recoil against the resonant sneutrino. We now take up
the study of these signals in detail.
\section{Signal Isolation}

\bigskip\noindent In this section we discuss various strategies for
identifying the signals that a sneutrino (of muonic or tauonic)
flavour has been produced in $e^+e^-$ interactions at a linear
collider and has decayed subsequently.  The numerical analysis has
been carried out for center-of-mass energy of 500~GeV for both the
cases, viz. the associated production of $\snu_\mu$ or of $\snu_\tau$
and their subsequent cascades to the four classes of
$R$-parity-violating signals listed above. We note that the analysis
at a 500~GeV linear collider provides sufficient physics insight into
isolating the $R$-parity-violating signal, and renders an analysis of
a 1~TeV machine, at this stage, redundant. Similarly, we have mostly
analysed the luminosity option ${\cal L} = 100$~fb$^{-1}$, since that
provides conservative estimates of statistical fluctuations in the SM
background.

\bigskip\noindent This section has been broken up into four
subsections for the four classes of signals listed above and we have
presented differential cross-sections for the parameters which show
the most significant deviations from the Standard Model
background. The latter involves the calculation of many diagrams, and
has been generated using the MadGraph package\cite{Stelzer:1994ta} and the
Madevent\cite{Maltoni:2002qb} Monte Carlo generator.

\subsection{The $e^+e^-\gamma$ final state}

\bigskip\noindent This final state arises from the direct $R$-parity
violating decay of the sneutrino into an $e^+e^-$ pair, with, of
course an associated photon from the initial state. The branching
ratio of the sneutrino to this mode is quite significant for 
 $\lambda_{1j1} \sim 0.03$ and hence the signal has a reasonable
cross-section. To detect this final state, we impose a 
set of acceptance sets, namely that each of the particles must not be 
too close to the beam pipe, 
\beq
\left| \eta (e^\pm) \right| \, , \, \left| \eta (\gamma) \right| < 2.0
      \label{eeg_cuts_rap}
\eeq
and that they should carry  sufficient transverse momenta
\beq
  p_T(e^\pm) > 10 \gev \, \quad {\rm and} \quad 
  p_T(\gamma) > 20 \gev \ .
      \label{eeg_cuts_pt}
\eeq
In addition, each pair of the final state particles should be well 
separated:
\beq
\delta R > 0.2 \ ,
      \label{eeg_cuts_dr}
\eeq
where $(\delta R)^2 \equiv (\Delta \phi)^2 + (\Delta \eta)^2$ with 
$\Delta \eta$  and $\Delta \phi$ respectively 
denoting the separation in rapidity and azimuthal angle. 
\begin{figure}[htb]
\vspace*{6ex}
\epsfxsize=6cm\epsfysize=6.0cm\epsfbox{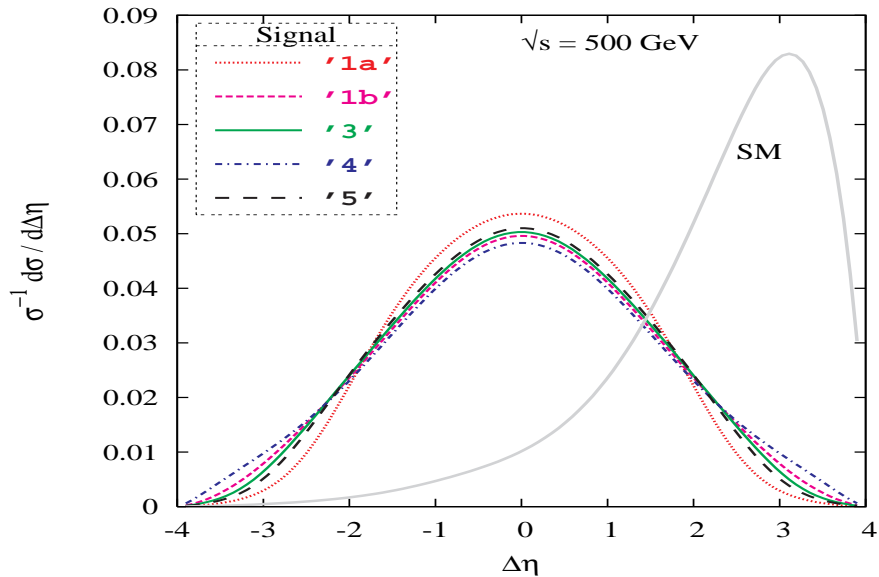} 
\caption{\em The normalized distribution in the difference 
        of electron and production rapidities for the 
	$e^+e^-\gamma$ final states at a $500~GeV$ linear collider, 
        and for the different Snowmass points. Also shown in 
        grey is the SM background. The cuts of eqn.
        (\ref{eeg_cuts_rap}--\ref{eeg_cuts_dr}) have been imposed.}
   \label{fig:delta_eta}
\end{figure}
\bigskip\noindent
Even with such cuts, the SM background, originating from radiative
Bhabha scattering, far overwhelms the signal, in fact almost by a
factor of 200. It is thus imperative to identify phase space variables
that would be preferentially sensitive to scalar production thereby
accentuating the signal to noise ratio.  An obvious such 
variable is the energy $E_\gamma$ of the recoil photon, which would be 
monochromatic in the case of the signal and a continuum for 
the background. However, before we consider $E_\gamma$, it is more useful to 
consider the difference between the fermion rapidities, namely
\beq
\Delta\eta_{ee} = \eta_{e^+} - \eta_{e^-}
\eeq
While the signal peaks at zero and is symmetric about it, the SM background 
is highly skewed towards positive $\Delta\eta_{ee}$ on account 
of the strong
$t$-channel photon contribution to (radiative) Bhabha scattering
(see Fig.\ref{fig:delta_eta}). 
Thus, if charge measurement of the electron and positron is 
straightforward and very efficient, requiring $\Delta \eta < 0$ 
would reduce the signal by only a factor of 2 while eliminating a very 
large part of the background. 
However, even if charge identification 
is not possible (or efficient), we could still consider 
$|\Delta\eta_{ee}|$, rather than $\Delta\eta_{ee}$
itself. Clearly, 
cutting-off higher values of $|\Delta\eta_{ee}|$ can reduce the
background considerably without significantly hurting the signal. A
detailed evaluation shows that
\begin{equation}
|\Delta\eta_{ee}| \leq 1.7
    \label{eeg_cuts_Deta}
\end{equation}
is the most suitable cut, i.e. the one which produces the largest
significance $N_{signal}/\sqrt{N_{SM}}$, where $N = \sigma
\cal{L}$. With this cut, the signal is reduced only by around 25\%
(with slight variations for different Snowmass points), while the
background is reduced by a factor larger than 4. We have, therefore,
implemented this cut in our subsequent numerical analysis of the
$\gamma e^+ e^-$ signal.

\begin{figure}[htb]
\begin{center}
\includegraphics[height=5.8in]{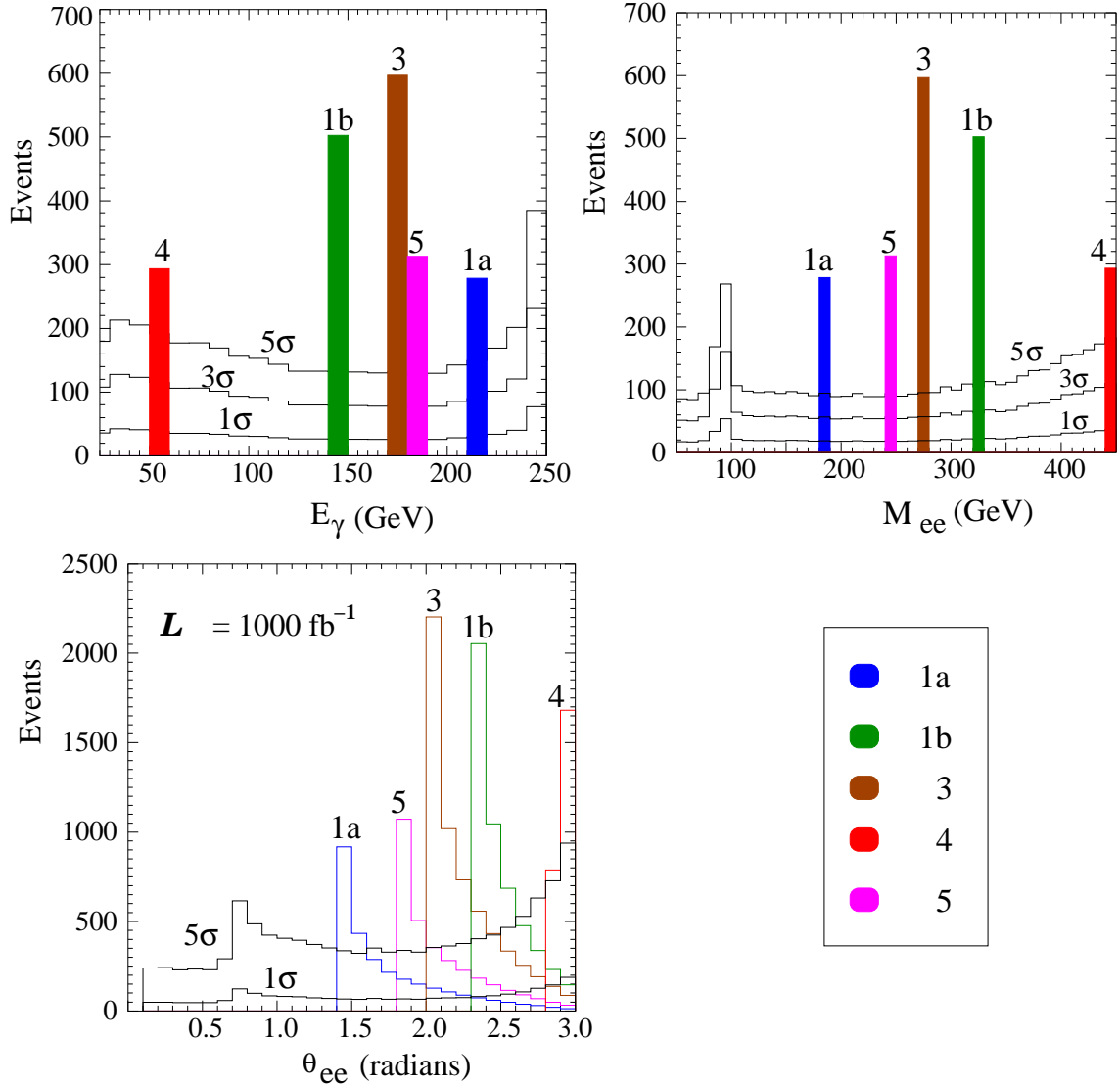}
\end{center}
\caption{\footnotesize\it
Illustrating kinematic distributions for the $e^+e^-\gamma$ final states at a
$500~GeV$ linear collider. Snowmass points are colour-coded and black lines
correspond to statistical fluctuations in the SM background. The cuts of eqn.
        (\ref{eeg_cuts_rap}--\ref{eeg_cuts_Deta}) have been imposed. 
	The first two
graphs correspond to luminosity $L = 100$~fb$^{-1}$.}
   \label{fig:eeg_distrib}
\end{figure}

\bigskip\noindent
We may now consider distributions in  ($a$) $E_\gamma$, which 
is our trigger, ($b$) the invariant mass
$M_{ee}$ of the $e^+e^-$ pair in the final state, which should peak at
the resonant mass, and ($c$) the opening angle $\theta_{ee}$ between
the $e^+e^-$ pair. 
While $E_\gamma$ and $M_{ee}$ are essentially the same observable in this 
case, we nonetheless include it as a counterpoint to the other cases 
to be discussed below.

\bigskip\noindent
In Figure \ref{fig:eeg_distrib} we show the signal distributions for the
case $\l_{121} = 0.03$ in the above variables with binnings which are
more-or-less consistent with the resolution(s) expected at a high
energy $e^+e^-$ collider, like, for example, Tesla. These are
represented by coloured histograms for the Snowmass points 1a, 1b, 3,
4 and 5. Rather than showing the large SM backgrounds, we have given
the statistical (Gaussian) fluctuations at 1, 3 and 5 standard
deviations. It is immediately apparent that the photon spectrum will
show clear peaks corresponding to recoil against a sneutrino. This
feature, expectedly, repeats itself in the invariant mass distribution.

\bigskip\noindent 
The opening angle between the $e^+e^-$ pair also
shows peaks tailing off towards large angles, but with a clear lower
cut-off depending on the sneutrino mass. It is clear
that the signal is somewhat less prominent, but still discernible,
when we consider this variable. This graph has been drawn assuming a
luminosity ${\cal L} = 1000$~fb$^{-1}$, unlike the previous ones,
which are for ${\cal L} = 100$~fb$^{-1}$.

\bigskip\noindent 
Finally, we should note that, for this particular
final state, the signal is extremely sensitive to the value of
$\l_{121}$. This is because both the sneutrino production
cross-section and the sneutrino branching ratio to an $e^+e^-$ pair
are proportional to $\l_{121}^2$. This quartic dependence ensures that
even a moderately lower value such as $\l_{121} = 0.01$ will ensure
that the signal is hardly discernible over the SM background even with
the high luminosity option ${\cal L} = 10^3$~fb$^{-1}$. The same
features repeat themselves for the $\l_{131}$ coupling. We therefore
turn to the other possible final states, which are related to more
robust decay modes of the sneutrino resonance.

\subsection{$e^+e^- \gamma \not{\!\!E_T}$ final states}

\noindent 
This final state differs from the last in having a
substantial amount of missing energy in addition to a trigger photon
and an $e^+e^-$ pair in the final state. A glance at Table~3 will
establish the fact that this channel corresponds to large branching
ratios both for the $\l_{121}$ and $\l_{131}$ cases. It arises (see
Table 2) from the $R$-parity-conserving decay of a sneutrino to a
same-flavour lepton and a neutralino, followed by three-body decay of
the neutralino through the $R$-parity violating coupling, with the
missing energy component coming from neutrinos in the final state.
Since the decay of the sneutrino to neutralinos is governed by gauge
couplings, this channel is suppressed only quadratically by lower
values of $\l_{1j1}$ --- and is hence considerably more robust than
the channel considered in the previous subsection.  Moreover, the SM
background, which comes from higher order processes than radiative
Bhabha scattering, is only at the level of about 36 events for ${\cal
L} = 10$~fb$^{-1}$, which is considerably below the signal, which is
in excess of a hundred events, as shown in Table 3. Thus, one can
expect an excess in the total cross-section over fluctuations in the
background even for $\l_{1j1}$ as low as 0.01. For the differential
cross-sections, the deviations are even more striking and hence much
lower values of the $\l_{1j1}$ coupling can be probed.

\begin{figure}[htb]
\begin{center}
\includegraphics[height=6in]{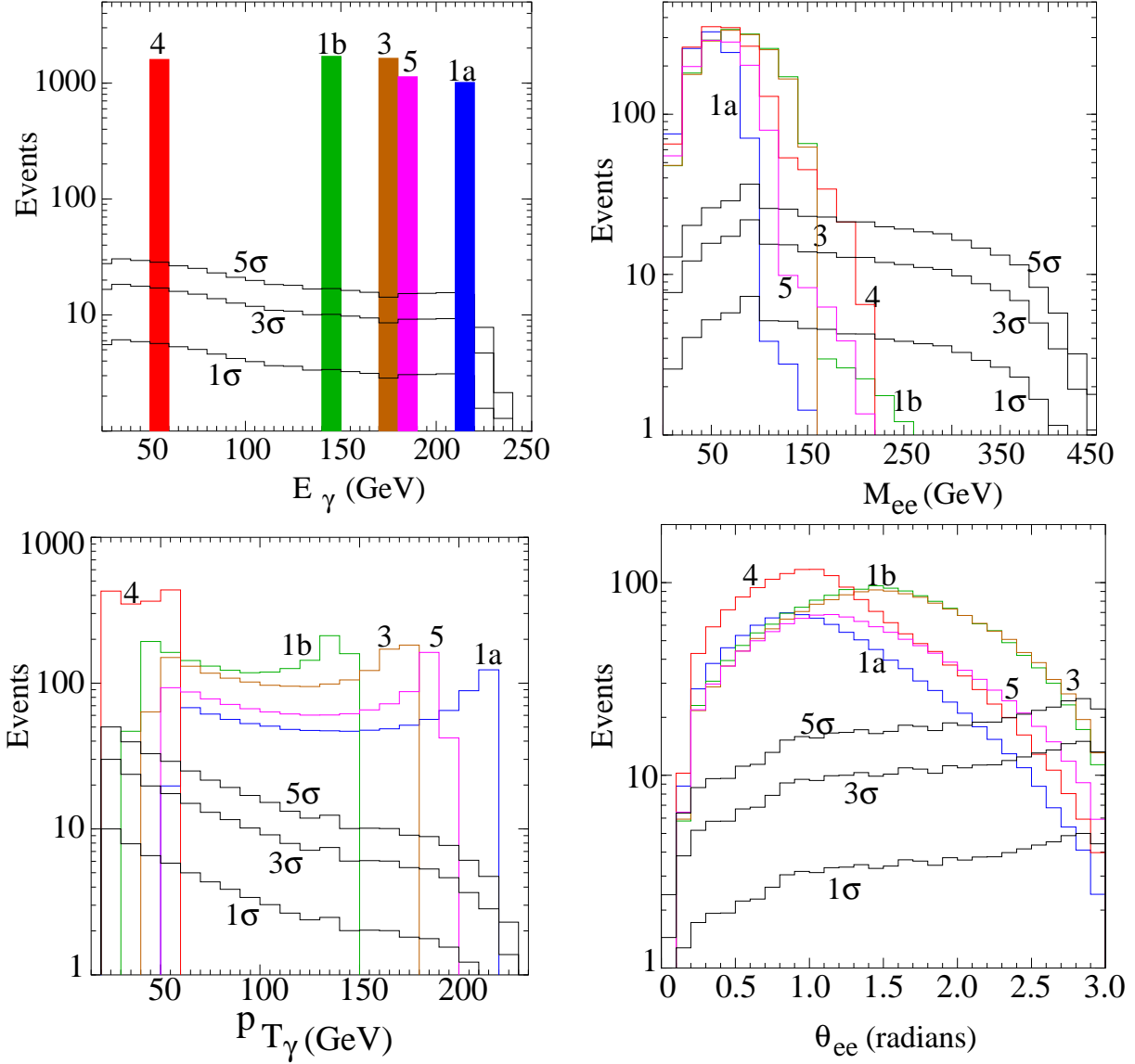}
\end{center}
\caption{\footnotesize\it
Illustrating the distributions in the excess in events over SM
predictions for the $e^+e^-\gamma \not{\!\!\!E_T}$ final states at a
$500~GeV$ linear collider.  The cuts of 
eqns.(\ref{eeg_cuts_rap}--\ref{eeg_cuts_dr}) and eqn.(\ref{cut_missing}) 
have been imposed. The colour coding is the same as
in {\rm Figure \protect\ref{fig:eeg_distrib}}. 
The RPV coupling involved here is $\lambda_{121}$
and the luminosity is ${\cal L} = 100$~fb$^{-1}$.}
   \label{fig:eegetsl_distrib}
\end{figure}

\noindent
Before proceeding further,  it is necessary to delineate the 
kinematic requirements that we seek to impose. As for the leptons 
and the photon, we choose the cuts to be the same as before, namely 
those listed in eqns.(\ref{eeg_cuts_rap}--\ref{eeg_cuts_dr}). In addition, 
we demand that the missing transverse momentum be 
sufficiently large, {\em viz.}
\beq
   \not{\!p}_T > 20 \gev
     \label{cut_missing}
\eeq
for it to be considered a genuine physics effect. 
Since the SM backgrounds
have a different source from the last case, it is not meaningful to
implement a cut on $|\Delta\eta_{ee}|$.

\bigskip\noindent 
The kinematic variables of interest for this channel
are similar, but somewhat different from the last case. As before, the
recoil photon spectrum ($E_\gamma$) should show a peak corresponding
to the resonance, and indeed it does, as a glance at 
Figure~\ref{fig:eegetsl_distrib} will
show. Since we have chosen $\l_{121} = 0.03$ for this graph, the peaks
are tall and sharp and cannot be missed by any means. In fact, these
are roughly two orders of magnitude above the $5\sigma$ background
fluctuation, which means that we will get observable effects even if
$\l_{121}$ is an order of magnitude smaller, say $\l_{121} \approx
0.003$.  

\bigskip
\noindent 
The other variables plotted in Figure~\ref{fig:eegetsl_distrib} are as
follows. The $e^+e^-$ invariant mass is no longer peaked at the
sneutrino mass since one of the $e^+$ and $e^-$ arises from three-body
decays of the neutralino. However, there is still a substantial
deviation from the background fluctuations. In fact, the second graph
in Figure~\ref{fig:eegetsl_distrib} shows that for low values of $\l_{121}$, 
the signal can be considerably enhanced by imposing a kinematic cut 
$M_{ee} < 250$~GeV. The third box shows the distribution in photon transverse
momentum, which, for smaller $\l_{121}$ will show modest deviations at
the right end. The last box shows the $e^+e^-$ opening angle
$\theta_{ee}$, which likewise shows deviations at the lower end. It is
worth mentioning that there is no significant deviation in the shape
of the missing energy and momentum curves, though, of course, there
will be an overall excess if $\l_{121}$ is large enough.

\bigskip\noindent We have not exhibited the curves for a $\l_{131}$ coupling
because they reproduce the same qualitative features, though the actual numerics
has slight differences.

\subsection{$e\mu \gamma \not{\!\!\!E_T}$ and $e\tau \gamma \not{\!\!\!E_T}$ final
states}

\noindent
The presence of an $R$-parity-violating coupling also ensures that
there will be significant numbers of sneutrinos which decay through
channels with a final state $\mu^\pm$ or a final state $\tau^\pm$ in
addition to a photon and an electron.  Such decay modes will also have
substantial missing energy from escaping neutrinos. Since the neutrino
flavours cannot be tagged, there will also be a substantial background
from SM processes with $W^+W^-$ pairs. If the $R$-parity-violating
coupling is $\l_{121}$ we can expect $e\mu$ combinations, while if the
coupling is $\l_{131}$ we can expect $e\tau$ combinations. However,
the cross-sections are not identical, since the cascade decays are not
the same.  This is due to the presence of low-lying $\stau$ states for
the Snowmass points under consideration.

\bigskip\noindent The analysis of these final states follows that of
the $\gamma e^+e^-\not{\!\!\!E_T}$ 
state quite closely. We impose precisely the same
kinematic cuts on the $\mu^\pm$ or $\tau^\pm$ as was imposed on the
electron and keep other cuts also the same. As in that case, the
signal cross-sections for $\l_{1j1} = 0.03$ are quite large and, in
fact, quite a few times larger than the background, which, for ${\cal
L} = 10$~fb$^{-1}$ is at the level of about 22 events for both $e\mu
\gamma \not{\!\!\!E_T}$ and $e\tau \gamma \not{\!\!\!E_T}$ final
states. Once again, the signal is expected to fall as $\l_{1j1}$
decreases, in which case it would be necessary to look at the
differential cross-sections. These, in turn, will resemble those of
Figure~\ref{fig:eegetsl_distrib} closely, because the actual
kinematics is very similar, all leptons appearing massless at the
energies under consideration. In the interests of brevity, we do not
exhibit the actual graphs, but merely note the following points:
\begin{itemize} 
\item The photon spectrum is, as usual, peaked at values
corresponding to the sneutrino mass. 
\item The $e\ell$ ($\ell = \mu, \tau$) invariant mass does not show
sharp peaks, but shows a kinematic boundary around $M_{e\ell} \simeq
\sqrt{s}/2$.
\item The transverse momentum of the photon peaks at high values around 
200~GeV. This feature distinguishes it from the background fluctuations,
which tend to fall uniformly as $p_{T\gamma}$ increases. The peaking, which
is very prominent in the figure shown, would becomes more modest if $\l_{1j1}$ 
were decreased. 
\item The $e\ell$ opening angle shows
modest (for smaller $\l_{1j1}$) peaking in the first quadrant. which again
deviates from the background, which prefers a back-to-back $e\ell$ pair.  
\item
The missing transverse momentum ($\not{\!\!p}_T$) distribution is
almost identical with that of the background. \end{itemize}

\noindent The most important feature of the $e\mu \gamma
\not{\!\!\!E_T}$ and $e\tau \gamma \not{\!\!\!E_T}$ final states is
that if a sneutrino is produced in sufficient numbers then one of the
two final states will exhibit an excess over the SM background {\it if
the $ee \gamma \not{\!\!\!E_T}$ state shows an excess}.  The other
will not, unless, indeed, both the $\l_{121}$ and $\l_{131}$ couplings
are present\footnote{There are strong constraints on the product
$\l_{121}\l_{131}$ from the non-observation of 
various decays forbidden in the Standard Model such as 
$\tau \to \mu\gamma$ or $\tau \to 3 e$~\protect\cite{fcnc}.  
It is, therefore, usual to set one of them to
zero. This is also consistent with our declared policy of considering
only one dominant coupling.}.  The existence of both a $ee \gamma
\not{\!\!\!E_T}$ and a $e\mu \gamma \not{\!\!\!E_T}$ signal is a
hallmark of a muonic sneutrino $\snu_\mu$, while the existence of both
a $ee \gamma \not{\!\!\!E_T}$ and a $e\tau \gamma \not{\!\!\!E_T}$
signal indicates production of a tau sneutrino $\snu_\tau$. Thus,
establishing the existence of the signal is the primary goal of the
analysis, and this, of course,  is facilitated by considering the
kinematic distributions discussed above.

\subsection{Multilepton final states}

\noindent 
Ten of the eighteen final states listed in Table 3 consist
of a hard photon and four identifiable leptons, of different
flavours. The number of events expected in these channels varies very
widely, as a glance at Table~3 will show.  Nevertheless, these
channels have practically no SM background, as a simple consideration
will show. We have already noted that the cross-section for producing
a photon and {\sl two} leptons is at the level of about 2~fb. To have
two more, we require to radiate a further gauge boson, which then
decays leptonically. This leads to suppression by at least the
electromagnetic coupling $\alpha$, i.e. by two orders of magnitude, 
provided, of course, that we assume minimum isolation criteria for
every pair of leptons. We
thus predict SM backgrounds at the level of 0.01~fb. By contrast, the
$R$-parity-violating signal is at the level of a femtobarn, which
means that it will stand out very clearly over the background.

\bigskip\noindent 
The presence of large numbers of leptons in the
final state is generally a signal for dileptons or of
$R$-parity-violating couplings, though the actual violation of lepton
number cannot be empirically established. This is because the missing
energy and momentum component could be due to an unknown number of
neutrinos carrying the necessary flavours to keep lepton number
conserved. Nevertheless, such explanations have been tried earlier,
whenever (seemingly) unexpected numbers of leptons have appeared in
the final state. For the present case, in addition to the presence of
four leptons (of which one is always an electron), we have a hard
associated photon, whose energy will peak at a value indicative of the
sneutrino resonance. The combination of such a mono-energetic photon
with multi-lepton final states, and with a cross-section at the
femtobarn level, would be difficult to explain away by any other
hypothesis than the present one.

\bigskip\noindent 
It is important, however, not to be too upbeat about
photon plus four-lepton final states as a signature of
$R$-parity-violating sneutrinos. This is because the cross-section for
such states depends heavily on the neutralino couplings and hence on
the point in the parameter space. For example, Table~3 shows that, for
a $\l_{121}$ coupling, there are no such signals even at the 0.1~fb
level for the Snowmass point 3. Absence of such signals, then, is not
unexpected, and should not be construed in a negative sense for the
model.

\bigskip\noindent In fact, of the above set of signals for sneutrino
decay, the most important points to note are ($a$) the mono-energetic
photon, and ($b$) the presence of one or more of the different final
states arising from sneutrino decay, which can be isolated from the
background by considering the distributions exhibited in
Figures~\ref{fig:eeg_distrib} and \ref{fig:eegetsl_distrib}. Though we
may not observe the full set of final states, we should certainly see
{\it something}, which would then clearly point to recoil of the
photon against a resonant particle with leptophilic couplings.

\section{Distinguishing a scalar from a vector resonance}

\bigskip\noindent 
As we have demonstrated in the previous section, it should be
possible, for a wide range of parameters, to establish a resonance by
triggering on a recoil photon of fixed energy and identifying a
variety of associated final states with leptons and missing
energy. However, it requires some more effort to identify the
resonance with a sneutrino of $R$-parity-violating supersymmetry.
The first step in such an exercise would be to determine the spin of
the resonance, a task that is best performed by analysing the angular
distributions.

\begin{figure}[htb]
\begin{center}
\includegraphics[height=3in]{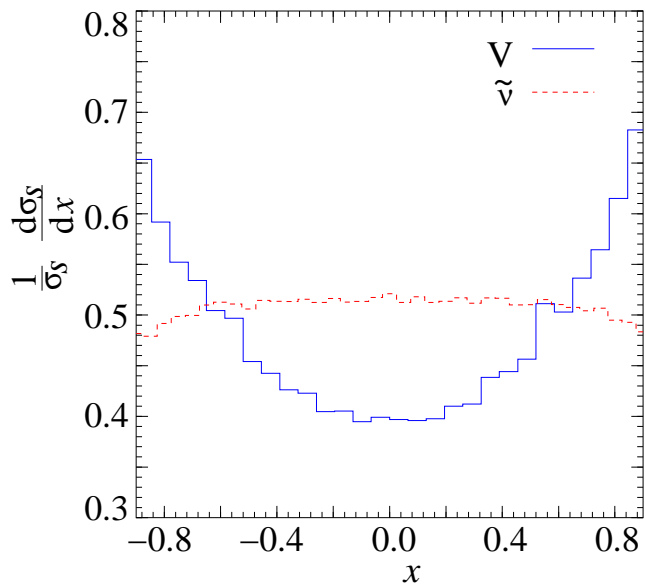}
\end{center}
\caption{\footnotesize\it Illustrating the
distribution in normalised differential cross-section of the signal
(excess over SM) with respect to $x=\cos\theta_e$. The dashed red 
(solid blue) curve shows the expectation for a scalar (vector) resonance.}
   \label{fig:distinguish}
\end{figure}

\bigskip\noindent
As far as the photon's distribution is concerned, it is primarily
driven by its $t$--($u$--)channel nature and hence is not very
sensitive to the spin of the resonance. Similarly, for the $\gamma
e^+e^-$ final state, it is not enough to construct just the angular
distribution of the electron/positron in the laboratory frame. For,
with the dilepton system recoiling against the photon, the effect of
the consequent boost tends to mask the smaller differences due to
spin.  Thus, the recoil needs to be corrected for, or, in other words,
we need to construct the angular distribution in the {\it rest frame}
of the $e^+e^-$ pair, i.e. of the resonance.  Denoting  the angle
between the final state $e^+/e^-$
and the parent resonance (whose direction is identical with the boost
axis, which in turn is opposite to that of the photon) by $\theta_e$, 
we construct
the distribution in $x \equiv \cos \theta_e$. This is exhibited in
Figure~\ref{fig:distinguish}, where the red~(blue) histograms
illustrate the normalised distribution expected for scalar~(vector)
particles.

\bigskip\noindent 
Even a cursory glance at the figure shows that there is a clear
difference between the more-or-less flat scalar distribution and the
vector distribution which shows a moderate depletion in the transverse
direction. In order to see if these would be actually observable, we
need to make an estimate of the possible errors in the histogram(s) of
Figure~\ref{fig:distinguish}. For a cross-section $\sigma_S \simeq
60$~fb and a luminosity of 100~fb$^{-1}$, the error in the normalised
cross-section comes out to be of the order of 0.01, which is clearly
much smaller than the actual difference between the two histograms. It
follows that one should be able to make a clear distinction between a
scalar and a vector resonance in $\gamma e^+e^-$ final states.

\bigskip\noindent 
The other final states considered in the previous
section, which contain a substantial missing energy component, are not
amenable to the reconstruction discussed above. Hence, for a small
$R$-parity-violating coupling, e.g. $\l_{1j1} \sim 0.01$, it may be
difficult to identify the scalar nature of the resonance with any
certainty. Of course, we always have the option of collecting higher
luminosity, in which case there will be a significant number of
$\gamma e^+e^-$ final states.

\bigskip\noindent 
A different method presents itself in the context of 
beam polarization. While we have, until now, considered only unpolarised
beams,  a high degree of beam polarisation is a realistic possibility at a high
energy linear $e^+e^-$ collider, and could be of considerable help in enhancing
the signal {\it vis-\'a-vis} background for many of the distributions shown
in the text. Furthermore, note that while the process 
$e^+e^- \to \gamma V$, with $V$ denoting a generic spin-one 
particle is enhanced if the electron and the positron 
have opposite helicities, the case of the (pseudo-)scalar $S$ prefers 
the helicities to be the same. 
In general, if $\eta_1$ and $\eta_2$ be the helicity states
of the initial $e^+e^-$ pair, then we have
\beq
\barr{rclcl}
\sigma(e^+e^- \to \gamma S) 
&\propto& (1 + \eta_1\eta_2) \, \left(|{\cal S}|^2 + |{\cal P}|^2 \right) \, 
         + 2 \, (\eta_1 + \eta_2) \, Re({\cal S \, P}^*) 
& \quad & {\rm for} \, \left({\cal S} + {\cal P} \, \gamma_5 \right)
 \\[1ex]
\sigma(e^+e^- \to \gamma V) 
&\propto& (1 - \eta_1\eta_2) \, (v^2 + a^2) - 2 \, v \, a \, (\eta_1 - \eta_2) 
& & {\rm for} \, \gamma_\mu \, (v + a \, \gamma_5) 
\earr
\eeq
It follows that ($a$) correct beam polarisation can enhance the signal
by a factor as large as 4 (assuming 100\% beam polarisation), while
the opposite polarisation can completely kill the signal, and ($b$)
the beam polarisation which is good for detecting a scalar resonance
is bad for detecting a vector resonance, and vice versa. In fact, this
is another way to distinguish between scalar and vector resonances, if
we can choose the beam polarisations at will. Even without maximal 
polarization, a study of the polarization-dependence can obviously 
shed much light on this issue. We, however, desist from discussing 
this any further as we feel that it
calls for a separate study in its own right.

\bigskip\noindent One final question remains. Assuming that we have
seen a scalar dilepton, how do we know if it is a sneutrino of
$R$-parity-violating supersymmetry, or a dilepton of some other model
(e.g. a composite dilepton)?  The answer lies in the observation, or
otherwise, of photon plus four-lepton final states, which, as we have
seen, arise principally from the decay of heavier gaugino states.  The
mere presence of such states is an indication that the underlying
model is supersymmetry. A detailed discussion of these is beyond the
scope of this work and, indeed, premature at this stage. We also sound
a note of caution that such four-lepton final states may not always be
observable, as pointed out in the last section.
\section{Summary and Conclusions}

\noindent 
An $e^+e^-$ collider, with its small backgrounds, could be an ideal
machine to discover supersymmetry without $R$-parity, especially the
lepton-number-violating variety. The presence of $LL\bar E$ operators
allows us to have sneutrino exchanges, which increases the total
four-fermion cross-sections.  When the coupling is too small for such
effects to show up, it is still possible, if the sneutrino is light
enough, to excite sneutrino ($\snu_\mu$ or $\snu_\tau$) resonances in
$e^+e^-$ collisions.  However, this would happen only if the machine
energy is tuned to the resonance, which is unlikely. We, therefore,
suggest a study of sneutrino production in association with a photon
radiated from the initial state, in which case the necessary spread in
energy is obtained and large resonant cross-sections result, even with
the ${\cal O}(\alpha)$ suppression. When we take into account the fact
that the sneutrino can decay through its $R$-parity-conserving (gauge)
couplings as well as its $R$-parity-violating $\l_{1j1}$ coupling,
four classes of final states result. All of these are characterised by
a mono-energetic photon, which corresponds to recoil against a
resonant sneutrino.

\bigskip\noindent 
The first of these, viz. $ee\gamma$, is present only
if the $\l_{1j1}$ is large enough and is perhaps the best signal for
sneutrino production. It is characterised, apart from a mono-energetic
photon, by strong peaks in the $e^+e^-$ invariant mass and opening
angle and modest excesses in the rapidity difference between $e^+$ and
$e^-$.  In this case, we can also reconstruct the final-state $e^\pm$
angular distribution in the $e^+e^-$ centre-of-mass frame and thereby
find a clear distinction between a scalar and a vector resonance with
identical decay modes. If found, this would serve to clinch the issue
of whether the resonance seen is indeed a sneutrino and not, for
example, a vector dilepton. Next, we have $e\ell\gamma \not{\!\!E}_T$
states, where $\ell = e, \mu, \tau$, of which one of the combinations
$\ell = e, \mu$ or $\ell = e, \tau$ is expected to show excesses over
the SM background, while the third will not. In both cases, we
predict, apart from a peak in the photon spectrum, a softer $e\ell$
invariant mass distribution than the SM and modest excesses in the
photon distribution at high transverse momentum and in low values of
the $e^+e^-$ opening angle. Finally, we have a mono-energetic photon
accompanied by four leptons, of various flavours, in different
combinations. These have very little SM background and, if the
cross-sections are large enough, would be practically (if not quite)
smoking gun signals of $R$-parity-violating supersymmetry.

\bigskip\noindent 
In this work, then, we have studied sneutrino
production in association with a recoil photon. We have shown that the
spread in energies induced by the photon radiation causes a `return to
the sneutrino peak' and enables a high energy $e^+e^-$ collider to
act, in a sense, as a sneutrino factory, if the supersymmetric model
does not conserve $R$-parity. Not only will this extend the search
range in the $\l_{1j1}$ parameter far beyond what a naive study of
excesses in dilepton final states could achieve. but different final
states will serve to establish the case for a sneutrino resonance and
to pin down the $R$-parity-violating coupling responsible for the
process(es). We have shown sample studies at the Snowmass points in
the $R$-parity-conserving sector of the parameter space. Though far
from exhaustive, these serve to illustrate our point, and are expected
to be a useful guide to experimental physicists searching for
supersymmetry when the high energy $e^+e^-$ collider is finally built
and commences operation.
\section*{Acknowledgements}
DC thanks the Department of Science and Technology, India for 
financial support through the Swarnajayanti Fellowship grant. 



\begin{thebibliography}{99}

\bibitem{Fayet:1977yc} P.~Fayet, {\it Phys.~Lett.} {\bf B69}, 489 (1977).

\bibitem{Farrar:1978xj} G.R.~Farrar and P.~Fayet, {\it Phys.~Lett.} 
{\bf B76}, 575 (1978).

\bibitem{Ibanez:1992pr} L.E.~Ibanez and G.G.~Ross, {\it Nucl.~Phys.}
{\bf B368}, 3 (1992).

\bibitem{Dreiner:vm} H.~Dreiner and G.G.~Ross, {\it Nucl.~Phys.}
{\bf B410}, 188 (1993).

\bibitem{nosusy} T.~Affolder \etal , {\it Phys.~Rev.~Lett.}
{\bf 88}, 04801 (2002); \\
B.~Abbott \etal (D0 Collab.), {\it Phys.~Rev.~Lett.} {\bf 83}, 4476 (1999); \\
F.~Abe \etal (CDF Collab.), {\it Phys.~Rev.~Lett.} {\bf 83}, 2133 (1999); \\
D.~Choudhury and S.~Raychaudhuri, {\it Phys.~Rev.} {\bf D56}, 1778 (1997).

\bibitem{HERA}
C.~Adloff \etal (H1 Collaboration), {\it Z.~Phys.} {\bf C74}, 191 (1997); \\
J.~Breitweg \etal (ZEUS Collaboration), {\it Z.~Phys.} {\bf C74}, 207
(1997).

\bibitem{Barger:1989rk}
V.~Barger, G.F.~Giudice and T.~Han, {\it Phys.~Rev.} {\bf D40}, 2987 (1989).

\bibitem{Bhattacharyya:1995pq}
G.~Bhattacharyya and D.~Choudhury,
{\it Mod.~Phys.~Lett.} {\bf A10}, 1699 (1995).

\bibitem{neutrino_mass}
Some studies are:\\
 R.M.~Godbole, P.~Roy and X.~Tata, {\it Nucl.~Phys.} {\bf B401}, 67 (1993);\\
 A.S.~Joshipura, V.~Ravindran and S.K.~Vempati, {\it Phys.~Lett.} 
{\bf B451}, 98 (1998);\\
M.~Drees, S.~Pakvasa, X.~Tata and T.~ter Veldhuis, {\it Phys.~Rev.} 
{\bf D57}, 5335 (1998);\\ 
E.J.~Chun and J.S.~Lee, {\it Phys.~Rev.} {\bf D60}, 075006 (1999);\\
A.~Abada and M.~Losada, {\it Nucl.~Phys.} {\bf B585}, 45 (2000);\\
A.S.~Joshipura, R.~Vaidya and S.K.~Vempati, {\it Phys.~Rev.} {\bf D65}, 
053018 (2002);\\
F.~Borzumati and J.S.~Lee, hep-ph/0207184.

\bibitem{bb0nu} H.~Klapdor-Kleingrothaus \etal, {\it Prog. Part. Nucl.
Phys.} {\bf 32}, 261 (1994); \\
J. W. F.~Valle, hep-ph/9509306; \\
G.~Bhattacharyya, H.~Klapdor-Kleingrothaus and H. Pas,
{\it Phys.~Lett.} {\bf B463}, 77 (1999).

\bibitem{rplimits}
For a summary of these limits, see for example, 
G.~Bhattacharyya, hep-ph/9709395;\\
B.C.~Allanach, A.~Dedes and H.K.~Dreiner, {\it Phys.~Rev.} {\bf D60}, 
075014 (1999); \\
M.~Chemtob, {\it Prog. Part. Nucl. Phys.} {\bf 54}, 71 (2005);\\
R.~Barbier \etal, hep-ph/0406039.


\bibitem{lspdecay} 
E.A.~Baltz and P.~Gondolo, {\it Phys.~Rev.} {\bf D57}, 2969 (1998);\\
H.~Dreiner, P.~Richardson and M.H.~Seymour,
{\it JHEP} {\bf 0004}, 008 (2000);\\
F.~Borzumati, R.M.~Godbole, J.L.~Kneur and F.~Takayama, hep-ph/0108244.

\bibitem{Dawson:1985vr}
S.~Dawson, {\it Nucl.~Phys.} {\bf B261}, 297 (1985).

\bibitem{Moreau:2000bs}
F.~Borzumati, J.L.~Kneur and N.~Polonsky, {\it Phys.~Rev.} {\bf D60}, 
115011 (1999);\\
G.~Moreau, E.~Perez and G.~Polesello, {\it  Nucl.~Phys.} {\bf B604}, 
3 (2001);\\
H.~Dreiner, P.~Richardson and M.H.~Seymour, {\it Phys.~Rev.} {\bf D63}, 
055008 (2001);\\
D.~Choudhury, S.~Majhi and V.~Ravindran, hep-ph/0207247.


\bibitem{DCSR} D.~Choudhury and S.~Raychaudhuri, hep-ph/9807373.

\bibitem{HERA_expl} 
D.~Choudhury and S.~Raychaudhuri, {\it Phys.~Lett.} {\bf B401}, 54 (1997); \\
G.~Altarelli \etal, {\it  Nucl.~Phys.} {\bf B506}, 3 (1997); \\
H.~Dreiner and P.~Morawitz, {\it  Nucl.~Phys.} {\bf B503}, 55 (1997); \\
J.~Kalinowski \etal, {\it Z. Phys.} {\bf C74}, 595 (1997); \\
T.~Kon and T.~Kobayashi, {\it Phys.~Lett.} {\bf B409}, 265 (1997);\\ 
G.~Altarelli, G.F.~Giudice, and M.L.~Mangano, {\it  Nucl.~Phys.} 
{\bf B506}, 29 (1997); \\
J.~Ellis, S.~Lola, and K.~Sridhar, {\it Phys.~Lett.} {\bf B408}, 
252 (1997); \\
M.~Carena, D.~Choudhury, S.~Raychaudhuri and C.E.M.~Wagner, 
{\it Phys.~Lett.} {\bf B414}, 92 (1997).
        
\bibitem{Ghosh:1997bm}
D.K.~Ghosh, S.~Raychaudhuri and K.~Sridhar, {\it Phys.~Lett.} {\bf B396}, 
177 (1997).

\bibitem{hanmargo}
T.~Han and M.B.~Magro, {\it Phys.~Lett.} {\bf B476}, 79 (2000);\\
K.J.~Abraham, K.~Whisnant, J.M.~Yang and B.L.~Young,
{\it Phys.~Rev.} {\bf D63}, 034011 (2001).

\bibitem{Bhattacharyya:1994yc}
G. Bhattacharyya, D. Choudhury and K. Sridhar,
{\it Phys.~Lett.} {\bf B349}, 118 (1995);\\
D.~Choudhury,  R.M.~Godbole and G.~Polesello, {\it JHEP} {\bf 0208}, 
004 (2002).

\bibitem{Kal} J.~Kalinowski, R.~R\"{u}ckl, H.~Spiesberger and P.M.~Zerwas,
{\it Phys.~Lett.} {\bf B414}, 297 (1997).

\bibitem{Hewett:1997ce}
J.L.~Hewett and T.G.~Rizzo, {\it Phys.~Rev.} {\bf D56}, 5709 (1997).

\bibitem{Hikasa:1999wy}
E.L.~Berger, B.W.~Harris and Z.~Sullivan, {\it Phys.~Rev.} {\bf D63}, 
115001 (2001);\\
K.~Hikasa, J.M.~Yang and B.~Young, {\it Phys.~Rev.} {\bf D60}, 114041 (1999);\\
D.~Choudhury, {\it Phys.~Lett.} {\bf B346}, 291 (1995);\\
A.~Datta, {\it Phys.~Rev.} {\bf D65}, 054019 (2002).

\bibitem{CCQR}M.~Carena, D.~Choudhury, S.~Lola and C.~Quigg, 
{\it Phys.~Rev.} {\bf D58}, 095003 (1998);\\
M.~Carena, D.~Choudhury, C.~Quigg and S.~Raychaudhuri, 
{\it Phys.~Rev.} {\bf D62}, 095010 (2000).

\bibitem{Feng:1996}
J.~Erler, J.L.~Feng and N.~Polonsky, {\it Phys.~Rev.~Lett.} {\bf 78}, 
3063 (1997); \\
J.L.~Feng, J.F.~Gunion and T.~Han,
{\it Phys. Rev.} {\bf D58}, 071701 (1998).

\bibitem{fcnc}K.~Agashe and  M.~Graesser, 
{\it Phys. Rev.} {\bf D54}, 4445 (1995); \\
D.~Choudhury and P.~Roy, {\it Phys.~Lett.} {\bf B378}, 153 (1996);\\
F.~Vissani and A.Yu.~Smirnov, {\it Phys.~Lett.} {\bf B380}, 317 (1996). 

\bibitem{Rai:2003}
S.K.~Rai and S.~Raychaudhuri, {\it JHEP} {\bf 0310}, 020 (2003).

\bibitem{Stelzer:1994ta}
T.~Stelzer and W.F.~Long,
{\it Comput.  Phys. Commun.} {\bf 81J}, 357 (1994).

\bibitem{Maltoni:2002qb}
F.~Maltoni and T.~Stelzer, {\it JHEP} {\bf 0302}, 027 (2003)
[hep-ph/0208156].

\end{thebibliography}
\end{document}